# Group Theoretical Analysis of Quasicrystallography from Projections of Higher Dimensional Lattices $B_n$

Mehmet Koca[a)] and Nazife Ozdes Koca [b)]

Department of Physics, College of Science, Sultan Qaboos University
P.O. Box 36, Al-Khoud, 123 Muscat, Sultanate of Oman

Ramazan Koc[c)]
Department of Physics, Gaziantep University, 27310, Gaziantep, Turkey

**Abstract**

A group theoretical discussion on the hypercubic lattice described by the affine Coxeter-Weyl group $W_a(B_n)$ has been presented. When the lattice is projected onto the Coxeter plane it is noted that the maximal dihedral subgroup $D_h$ of $W(B_n)$ with $h=2n$ representing the Coxeter number describes the $h$-fold symmetric quasicrystallography. Higher dimensional cubic lattices are explicitly constructed for $n = 4, 5, 6$. Their rank 3 Coxeter subgroups and maximal dihedral subgroups are identified. It has been explicitly shown that when their Voronoi cells are decomposed under the respective rank 3 subgroups $W(A_3)$, $W(H_2) \times W(A_1)$ and $W(H_3)$ one obtains the rhombic dodecahedron, rhombic icosahedron and rhombic triacontahedron respectively. Projection of the lattice $B_4$ onto the Coxeter plane represents quasicrystal structures with 8-fold symmetry. The $B_5$ lattice is used to describe the quasicrystals with both 5-fold and 10-fold symmetries. The lattice $B_6$ can describe a 12-fold symmetric quasicrystal as well as a 3D icosahedral quasicrystal depending on the choice of subspace of projections. The novel structures from the projected sets of lattice points are compatible with the available experimental data.

[a)] electronic-mail: kocam@squ.edu.om
[b)] electronic-mail: nazife@squ.edu.om
[c)] electronic-mail: koc@gantep.edu.tr

## 1. Introduction

Quasicrystallography gained enormous impetuous after the first discovery of the icosahedral quasicrystal by D. Shechtman [1]. Its point symmetry can be described by the Coxeter group $W(H_3)$ representing the icosahedral symmetry of order 120. Recent developments indicate that the quasicrystals exhibit 5-fold, 8-fold, 10-fold, 12-fold, and 18-fold symmetries. For a general exposition we refer the reader to the references on quasicrystallography [2-6]. In a recent paper [7] it was reported that a quasicrystallographic structure with 36-fold symmetry is possible. These predictions imply that no limitation exists on the order of the planar point symmetry of the quasicrystallography described by the dihedral groups. That reminds us the classification of the Coxeter-Weyl groups with different Coxeter numbers $h$ [8-11]. Every Coxeter-Weyl group has a dihedral subgroup $D_h$ of order $2h$. This paper will attempt to illustrate the relations between the group theoretical structures of the affine Coxeter groups $W_a(B_n)$ and the $h$-fold symmetric quasicrystallography obtained from higher dimensional cubic lattices by orthogonal projections. A general projection technique of the higher dimensional cubic lattice is prescribed by Duneau and Katz [12] but with no detailed group theoretical discussion on the symmetries of the lattices.

The Lie groups derived from the root systems of the Coxeter-Weyl groups are well known by the high energy physicists. Predictions of the standard model of the High Energy Physics described by the Lie group $SU(3) \times SU(2) \times U(1)$ [13-15] are heavily based on the Coxeter-Weyl group $W(A_2) \times W(A_1)$. The skeletons of the Grand Unified theories, $SU(5) \approx E_4$ [16], $SO(10) \approx E_5$ [17] and the exceptional group $E_6$ [18] are the respective Coxeter-Weyl groups $W(A_4)$, $W(D_5)$, and $W(E_6)$. It is expected that some of the Coxeter-Weyl groups with a Coxeter number $h$ may also play an important role in the study of the quasicrystallography. The argument is based on the following line of thought.

Any Coxeter group with a Coxeter number $h$ has a maximal dihedral subgroup $D_h$ of order $2h$ which acts in a certain Coxeter plane. (It is unfortunate that the same notation is also used for the Coxeter-Weyl group $W(D_n)$). In two recent papers [19-20] we have proposed that any quasicrystallographic structure with $h$-fold symmetry can be determined by projections of the higher dimensional lattices onto the relevant Coxeter planes. The Coxeter groups are naturally characterized by some integers known as Coxeter exponents [8-11]. In this paper we study the general structure of the root lattice of the affine Coxeter group $W_a(B_n)$. It is the simple cubic lattice in $nD$ Euclidean space with the point symmetry determined by the Coxeter-Weyl group $W(B_n)$ of order $2^n n!$. The projection of the 5D cubic lattice onto a plane and the projection of 6D cubic lattice into a 3D subspace have been studied earlier without using Coxeter group techniques [12, 21]. Projections of some 4D root lattices have been also studied earlier [22-24].

The group $W(B_n)$ itself can be regarded as an extension of the elementary abelian group $2^n$ by its permutation group $S_n$. The paper is organized as follows. In Section 2 we study the general structure of the Coxeter-Weyl group $W(B_n)$ with some emphasis on its maximal subgroups which could be useful for the projections of the hypercubic lattices. Section 3 deals with the study of the rank 3 subgroups of the Coxeter-Weyl groups $W(B_4)$, $W(B_5)$, and $W(B_6)$ and projections of some of their polytopes into 3D Euclidean spaces with different residual symmetries. The projection of the Voronoi cell of a higher dimensional lattice plays a crucial role in the description of the quasicrystallographic structures in 3D and 2D. Its structure for the $W(B_n)$ lattices will be pointed out. Section 4 is devoted to the projection techniques of the lattices onto the Coxeter planes and the projection of the 6D cubic lattice into 3D subspace with icosahedral symmetry. Our predictions are compared with experimental models in Section 5 and some conclusive remarks are added.

## 2. The Coxeter-Weyl group $W(B_n)$ and its Maximal Subgroups

The classification of the Coxeter-Weyl groups is well known [8-11]. It includes an infinite series of crystallographic groups $A_n$, $B_n$, $C_n$, $D_n$, and a finite number of crystallographic exceptional groups $G_2$, $F_4$, $E_6$, $E_7$, $E_8$. In addition to the above crystallographic groups there are an infinite number of noncrystallographic dihedral Coxeter groups $I_2(h)$ ($h \geq 5$, $h \neq 6$) and the two rank-3 and rank-4 noncrystallographic Coxeter groups $W(H_3)$ and $W(H_4)$ respectively. In this section we will be interested in the group theoretical structures of the Coxeter-Weyl group $W(B_n)$ of rank $n$. It is represented by the Coxeter-Dynkin diagram shown in Figure 1.

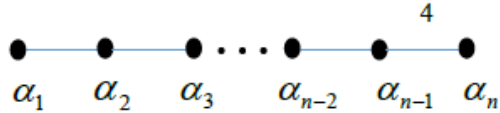

**Figure 1**: Coxeter-Dynkin diagram of the Coxeter-Weyl group $W(B_n)$

From left to right the nodes denote the long simple roots $\alpha_i$ ($i = 1, 2, ..., n-1$) with norm $\sqrt{2}$ and the last simple root $\alpha_n$ is the short root with norm 1. They represent linearly independent vectors in $n$-dimensional Euclidean space. Angle between any two adjacent simple roots with norm $\sqrt{2}$ is $120^0$ and the angle between the last two simple roots is $135^0$. Any two roots, not adjacent, are orthogonal to each other. The nodes also denote the reflection generators $r_i$ whose action on an arbitrary vector $\Lambda$ in the $n$-dimensional Euclidean space is given by

$$r_i \Lambda = \Lambda - \frac{2(\Lambda, \alpha_i)}{(\alpha_i, \alpha_i)} \alpha_i . \qquad (1)$$

It is useful to work in the dual space of the root space represented by the weight vectors $\omega_i$ defined by $(\omega_i, \frac{2\alpha_j}{(\alpha_j,\alpha_j)}) = \delta_{ij}$. When the direct space is associated with the root space the reciprocal lattice is associated with the weight space. The Cartan matrix of the root space (Gram matrix) and the metric tensor in the dual space are defined respectively by

$$A_{ij} = \frac{2(\alpha_i,\alpha_j)}{(\alpha_j,\alpha_j)}, \quad G_{ij} = (\omega_i,\omega_j) = (A^{-1})_{ij}\frac{(\alpha_j,\alpha_j)}{2}. \tag{2}$$

Let $l_i (i=1,2,...,n)$ be the set of orthonormal vectors $(l_i, l_j) = \delta_{ij}$ in $n$-dimensional Euclidean space. The simple roots of $W(B_n)$ can be written as

$$\alpha_1 = l_1 - l_2, \alpha_2 = l_2 - l_3, ..., \alpha_{n-1} = l_{n-1} - l_n, \alpha_n = l_n. \tag{3}$$

The root system consists of two sets, one set with $2n$ short roots $\pm l_i$ and the other with $2n(n-1)$ long roots $\pm l_i \pm l_j (i \neq j)$. Reflection generators $r_i$ act on the unit vectors as follows

$$r_1 : l_1 \leftrightarrow l_2, r_2 : l_2 \leftrightarrow l_3, ..., r_{n-1} : l_{n-1} \leftrightarrow l_n, r_n : l_n \to -l_n. \tag{4}$$

The generators $r_i$ generate the Coxeter-Weyl group $W(B_n)$. Equation (4) implies that the reflection generators leave the other unit vectors (not shown) invariant. The weight vectors can be determined from $\omega_i = (A^{-1})_{ij}\alpha_j$ as

$$\omega_1 = l_1, \omega_2 = l_1 + l_2, ..., \omega_{n-1} = l_1 + l_2 + \cdots + l_{n-1}, \omega_n = \frac{1}{2}(l_1 + l_2 + \cdots + l_n). \tag{5}$$

Any highest weight vector [25] in the weight space can be written as $\Lambda = a_1\omega_1 + a_2\omega_2 + \cdots + a_n\omega_n \equiv (a_1, a_2, ..., a_n)$ with integer coefficients $a_i \geq 0$. We will delete the commas between the integers as long as an integer does not exceed 9. An orbit of $\Lambda$ under the Coxeter-Weyl group $W(B_n)$ will be denoted by $W(B_n)\Lambda \equiv (a_1 a_2 ... a_n)_{B_n}$. With this notation, e.g., the orbits $(100...0)_{B_n} = \pm l_i$ and $(010...0)_{B_n} = \pm l_i \pm l_j, (i \neq j)$ represent the sets of short roots and long roots respectively. The orbit $(00...01)_{B_n} = \frac{1}{2}(\pm l_1 \pm l_2 \pm \cdots \pm l_n)$ represents the vertices of a cube in $n$-dimensional Euclidean space. Before we discuss the maximal subgroups of the group $W(B_n)$ we point out that the same Coxeter-Weyl group defines another Coxeter-Dynkin diagram $C_n$ where the short and long roots are represented by the roots $\alpha_1 = l_1 - l_2, \alpha_2 = l_2 - l_3, ..., \alpha_{n-1} = l_{n-1} - l_n, \alpha_n = 2l_n$. Here the short and long roots of $W(B_n)$ are interchanged but the symmetry group is the same as defined in (4). Certain maximal subgroups of $W(B_n)$ can be useful in the study of quasicrystals. One of the maximal subgroup of $W(B_n)$ is the Coxeter-Weyl group $W(D_n)$ with the Coxeter-Dynkin diagram given in Figure 2. It consists of only long roots of norm $\sqrt{2}$.

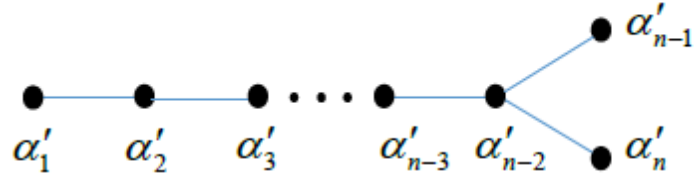

**Figure 2:** Coxeter-Dynkin diagram of $W(D_n)$

The group $W(D_n)$ is a maximal subgroup with an order of $2^{n-1}n!$ in the group $W(B_n)$ with an index 2. Denote by $r_i'$ the reflection generators of $W(D_n)$ and the simple roots by

$$\alpha_1' = l_1 - l_2, \alpha_2' = l_2 - l_3, ..., \alpha_{n-1}' = l_{n-1} - l_n, \alpha_n' = l_{n-1} + l_n. \tag{6}$$

The first $n$-1 simple roots are identical to the simple roots of $W(B_n)$ except the last one. The generators of $W(D_n)$ transform the unit vectors as in (4) but the last generator differs in its action

$$r_1': l_1 \leftrightarrow l_2, r_2': l_2 \leftrightarrow l_3, ..., r_{n-1}': l_{n-1} \leftrightarrow l_n, r_n': l_{n-1} \leftrightarrow -l_n, \tag{7}$$

This implies that one can identify $r_i' = r_i$, for $(i = 1, 2, ..., n-1)$ and we note that $r_n' = r_n r_{n-1} r_n$. The Coxeter-Dynkin diagram of $W(D_n)$ has a diagram symmetry $\gamma: \alpha_{n-1}' \leftrightarrow \alpha_n'$ which transforms $\gamma: l_n \rightarrow -l_n$ leaving the other unit vectors invariant and it can be identified with $r_n$. Therefore the automorphism group of $D_n$ is $Aut(D_n) \approx W(D_n):\mathbf{Z}_2 \approx W(B_n)$ where the group $\mathbf{Z}_2$ is generated by $\gamma$. We note that there is one exception to this general result: the automorphism group of $D_4$ is larger than $W(B_4)$ and it is the group $Aut(D_4) \approx W(F_4)$. We also note that some of the orbits of $W(D_n)$ and the orbits of $W(B_n)$ have identical number of vertices

$$\left|(10...0)_{D_n}\right| = \left|(10...0)_{B_n}\right|, \left|(010...0)_{D_n}\right| = \left|(010...0)_{B_n}\right|,$$
$$,..., \left|(00...1000)_{D_n}\right| = \left|(00...1000)_{B_n}\right| \tag{8}$$

The orbit $(00...01)_{B_n}$ is the union of two orbits of $W(D_n)$,
$$(00...1)_{B_n} = (00...10)_{D_n} \cup (00...01)_{D_n}. \tag{9}$$

The group $W(B_n)$ admits the group $W(B_{n-1})$ and $W(A_{n-1})$ as maximal subgroups. One of the interesting maximal subgroup of $W(B_n)$ is the dihedral group $W(I_2(h)) \approx D_h$ of order $2h$. Dihedral group $D_h$ plays an important role in the study of 2D quasicrystals with $h-fold$ symmetry. The argument goes as follows. The simple roots of $W(B_n)$ decompose in such a way that the corresponding reflection generators $r_1, r_3, ..., r_{n-1}$ commute pairwise as well as do the set $r_2, r_4, ..., r_n$ for even $n$. A similar decomposition can be done for odd $n$. Define now the generators $R_1$ and $R_2$ by $R_1 = r_1 r_3 ... r_{n-1}$ and $R_2 = r_2 r_4 ... r_n$ [26]. It is easy to show that the generators $R_1$ and $R_2$ act as reflections on the simple roots of the Coxeter diagram $I_2(h)$ and the Coxeter element $R_1 R_2$ represents a rotation of order $h$ in the plane spanned by the simple roots [27] of the Coxeter-Dynkin diagram $I_2(h)$. The integers $m_i$ of the Coxeter exponents $m_i \frac{\pi}{h} (i = 1, 2, ..., n)$ with $m_i = 1, 3, 5, ..., 2n-1$ and the Coxeter number $h = 2n$ [8-11] are useful in determination of the Coxeter plane in which one can have quasicrystallographic point symmetry of the projected set of points. The eigenvalues of the Cartan matrix in (2) can be written simply $\lambda_i = 2[1 - \cos(m_i \frac{\pi}{h})]$ [8]. To study the quasicrystallography the plane determined by the eigenvectors corresponding to the pair of eigenvalues $2[1 - \cos(m_i \frac{\pi}{h})]$, $2[1 - \cos((h - m_i) \frac{\pi}{h})]$ is of interest. One can choose a convenient set of orthogonal unit vectors obtained from the eigenvectors of the Cartan matrix [19-20]

$$\hat{x}_i = \frac{1}{\sqrt{h \lambda_i}} \sum_j \frac{2\alpha_j}{(\alpha_j, \alpha_j)} X_{ji} \tag{10}$$

where $\vec{X}_i$ is the eigenvector of the Cartan matrix given in (2) corresponding to the eigenvalue $\lambda_i$ with normalization where the last components of the eigenvectors all equal 1. The simple roots $\beta_i, \beta_{n+1-i}$ of the dihedral group $I_2(\frac{h}{m_i})$,

$$\begin{aligned} \beta_i &= \sqrt{2}[\sin(\frac{m_i \pi}{2h})\hat{x}_i + \cos(\frac{m_i \pi}{2h})\hat{x}_{n+1-i}], \\ \beta_{n+1-i} &= \sqrt{2}[\sin(\frac{m_i \pi}{2h})\hat{x}_i - \cos(\frac{m_i \pi}{2h})\hat{x}_{n+1-i}], (i = 1, 2, ..., \frac{n}{2}) \end{aligned} \tag{11}$$

determine the Coxeter plane in which the generators $R_1$ and $R_2$ act like reflection generators on the simple roots $\beta_i, \beta_{n+1-i}$.

The root lattice of $W(B_n)$ is the simple cubic lattice which is invariant under the affine Coxeter group $W_a(B_n)$ that can be generated by adding a generator $r_0$ to the set of generators of the group $W(B_n)$. The generator $r_0$ represents a reflection with respect to the hyperplane bisecting the highest long root $\tilde{\alpha} = \omega_2 = l_1 + l_2$. Its action on an arbitrary vector is a translation $r_0 : \Lambda \to \Lambda + \tilde{\alpha}$. A general vector of the root lattice then will be given by $p = b_1 \alpha_1 + b_2 \alpha_2 + \cdots + b_n \alpha_n, b_i \in \mathbf{Z}$ which can also be written in the weight space as $p = a_1 \omega_1 + a_2 \omega_2 + \cdots + 2a_n \omega_n, a_i \in \mathbf{Z}$. This indicates that a general vector of the lattice is the linear combinations of the unit vectors $l_i$ with integer coefficients. This also implies that the root lattice of $W(B_n)$ is generated by its short roots. The primitive cell of the lattice can be chosen as the cube with the vertices

$$0, l_i, l_i + l_j \ (i<j), l_i + l_j + l_k \ (i<j<k), ..., l_i + l_j + \cdots + l_k \ (i<j<...<k). \tag{12}$$

There are $2^n$ such cubes sharing the origin as a vertex. The Voronoi cells around the lattice points are congruent polytopes tiling the $n$-dimensional Euclidean space. The Voronoi cells of lattices are important in the theory of coding [28]. We denote the Voronoi cell around the origin by $V(0)$ and its structure is important for the canonical projection (strip projection, or cut and project technique) of the lattice points onto the Coxeter plane. Vertices of the Voronoi polytope $V(0)$ can be determined as the intersection of the hyperplanes surrounding the origin. They are the hyperplanes determined as the orbits of the fundamental weights

$$\frac{\omega_1}{2} = \frac{l_1}{2}, \frac{\omega_2}{2} = \frac{l_1 + l_2}{2}, \cdots, \frac{\omega_{n-1}}{2} = \frac{l_1 + l_2 + \cdots l_{n-1}}{2}, \omega_n = \frac{1}{2}(l_1 + l_2 + \cdots + l_n). \tag{13}$$

The Voronoi polytope $V(0)$ is then a cube around the origin with the vertices [29-30]

$$(00...01)_{B_n} = \frac{1}{2}(\pm l_1 \pm l_2 \cdots \pm l_n). \tag{14}$$

Before we proceed further we emphasize here that the lattices generated by the affine Coxeter-Weyl group $W_a(C_n)$ are identical to the root and weight lattices of the Coxeter-Weyl group $W_a(D_n)$ [28] since the short roots of $W(C_n)$ are identical to the root system of the group $W_a(D_n)$. The group $Aut(D_n) \approx W(B_n)$ should be taken into account when the point symmetry of the two lattices is of concern. Therefore in an $n$-dimensional Euclidean space with $n>3$ one can construct five different lattices with affine Coxeter-Weyl groups, two for $W_a(A_n)$, two for $W_a(D_n)$, and one simple cubic lattice described by $W_a(B_n)$. Of course if $n$ coincides with the rank of the exceptional groups the number of lattice will be more than 5. We have to determine the components of a lattice vector in the Coxeter planes defined by the pairs of unit vectors $(\hat{x}_1, \hat{x}_n)$, $(\hat{x}_2, \hat{x}_{n-1})$,.... Note that for odd $n$ one of the unit vectors is unpaired which represents the direction orthogonal to all Coxeter planes. The representation of the generators $R_1$ and $R_2$ of the dihedral subgroup can be put into block-diagonal matrices with $2\times 2$ and/or $1\times 1$ matrix entries. Some of the planes may display the crystallographic symmetries rather than the quasicrystallographic symmetries depending on the values of the Coxeter exponents $m_i$. The component of the simple cubic lattice vector in the basis of $\hat{x}_i$ is given by

$$p_i = \frac{1}{\sqrt{h\lambda_i}}(\sum_{j=1}^{n-1} a_j X_{ji} + 2a_n), \ a_i \in \mathbf{Z}. \tag{15}$$

This preliminary introduction on the hypercubic lattice and its symmetry group will be useful in the following chapters where we study the 8-fold, 5-fold, 10-fold, and 12-fold symmetric quasicrystal structures induced by the projections of the lattices $B_4, B_5,$ and $B_6$.

## 3. The Coxeter-Weyl groups $W(B_4)$, $W(B_5)$, $W(B_6)$ and projections of their polytopes into 3D subspaces

The projected copies, in 3D space, of the Voronoi cells of the root lattices of the Coxeter-Weyl groups $W(B_4)$, $W(B_5)$, and $W(B_6)$ are the rhombic dodecahedron, rhombic icosahedron, and rhombic triacontahedron respectively. This is quite well known in the literature but has never been presented in a systematic way. A similar work, not in the context of the affine Coxeter groups, has been studied by Kramer [31].

### 3.1 The Coxeter-Weyl group $W(B_4)$ and projection of its fundamental polytopes into a 3D subspace with octahedral symmetry

In the study of representations of the Lie groups the weight vectors $\omega_i \, (i=1,2,...,n)$ play an important role and are called fundamental weights. Hereafter we will call those polytopes obtained as the orbits of the fundamental weights as the fundamental polytopes. The vertices of the fundamental polytopes of the group $W(B_4)$ are given by

$$(1000)_{B_4} = \pm l_i,$$
$$(0100)_{B_4} = \pm l_i \pm l_j \, (i \neq j),$$
$$(0010)_{B_4} = \pm l_i \pm l_j \pm l_k \, (i \neq j \neq k), \qquad (16)$$
$$(0001)_{B_4} = \frac{1}{2}(\pm l_1 \pm l_2 \pm l_3 \pm l_4), \; i,j,k = 1,2,3,4.$$

These are well known 4D polytopes with 4D cubic symmetry. The first one represents the short roots of $W(B_4)$ which constitutes a 4D octahedron with 8 vertices whose facets (3-faces) are tetrahedra. The second polytope consists of 24 long roots as vertices and is known as the 24-cell with octahedral facets. Its full symmetry is the Coxeter-Weyl group $W(F_4)$ which embeds $W(B_4)$ as a subgroup with index 3. The third polytope with 32 vertices has two types of facets: tetrahedra and truncated octahedra. The last one is the 4D cube with 16 vertices consisting of the cubic facets.

The subspace we are interested in here is a 3D Euclidean space with the octahedral symmetry represented by the Coxeter-Weyl group $W(B_3)$ of order 48. Let us recall the isomorphism $W(B_3) \approx Aut(D_3) \approx Aut(A_3) = W(A_3):\mathbf{Z}_2$. Here one can choose the tetrahedral symmetry as the Coxeter-Weyl group $W(A_3) = \langle r_1, r_2, r_3 \rangle$. The Dynkin diagram symmetry $\mathbf{Z}_2$ generated by $\gamma$ which permutes the unit vectors as $l_1 \leftrightarrow l_3$ and $l_2 \leftrightarrow l_4$ extends the group to the octahedral symmetry. Using $\gamma$ and (4) for the actions of the generators $r_1, r_2$, and $r_3$ it is clear that the vector $\frac{1}{2}(l_1 + l_2 + l_3 + l_4)$ is invariant under the octahedral group. To describe the 3D Euclidean space where the octahedral group acts as the symmetry group, it is convenient to introduce a new set of orthonormal vectors defined by

$$t_0 = \frac{1}{2}(l_1 + l_2 + l_3 + l_4),\ t_1 = \frac{1}{2}(l_1 - l_2 + l_3 - l_4),$$
$$t_2 = \frac{1}{2}(-l_1 + l_2 + l_3 - l_4),\ t_3 = \frac{1}{2}(l_1 + l_2 - l_3 - l_4). \tag{17}$$

When the vectors in (16) are expressed in terms of the vectors $t_0, t_1, t_2$, and $t_3$ it will be simpler to identify the 3D vertices of the 4D polytopes. The 4D octahedron projects onto a cube in 3D with the vertices $\frac{1}{2}(\pm t_1 \pm t_2 \pm t_3)$. The 24-cell projected into the 3D space represents two octahedra as well as one cuboctahedron. The polytope with 32 vertices is projected into one cube and two truncated tetrahedra. Under the octahedral symmetry the union of two truncated tetrahedra forms a non-regular polyhedron with 24 vertices. Its faces are made of equilateral triangles, squares and rectangles.

The polytope $(0001)_{B_4}$ is more interesting since it constitutes the Voronoi cell of the 4D cubic lattice. Two vectors $\pm \frac{1}{2}(l_1 + l_2 + l_3 + l_4)$ from the set are projected to the origin. When expressed in terms of the unit vectors $t_0, t_1, t_2$, and $t_3$ and the component of an arbitrary vector along $t_0$ is deleted they will decompose under the octahedral group as two orbits, one with the vertices $(\pm t_1, \pm t_2, \pm t_3)$ representing the vertices of an octahedron and the other with vertices $\frac{1}{2}(\pm t_1 \pm t_2 \pm t_3)$ representing a cube. Union of these two orbits represents a rhombic dodecahedron [32] with 14 vertices as shown in Figure 3.

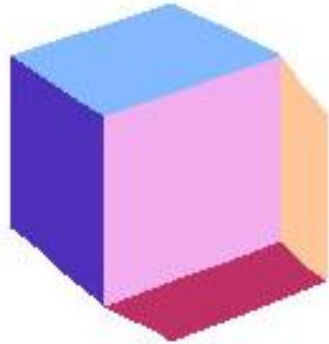

**Figure 3:** The rhombic dodecahedron

## 3.2 The Coxeter-Weyl group $W(B_5)$ and projection of its fundamental polytopes into 3D subspace with $W(H_2) \times C_2 \approx D_{5d}$ symmetry

One of the fundamental orbits of the group $W(B_5)$ is the polytope $(10000)_{B_5} = \pm l_i \, (i=1,2,...,5)$ which represents an octahedron in 5D Euclidean space whose facets are 5-cells (4-simplexes). We will see that when it is projected into a 3D space it represents a pentagonal antiprism where the subgroup $W(H_2) \times C_2 \approx D_{5d}$ acts as a point group. To understand this better we have to define a new set of orthonormal vectors $\hat{x}_i \, (i=1,2,...,5)$. The first four unit vectors $\hat{x}_i \, (i=1,2,...,4)$ is obtained by using the eigenvectors of the Cartan matrix of $W(A_4)$ and the fifth vector is chosen to be orthogonal to the rest [33]

$$\hat{x}_1 = \frac{1}{\sqrt{2(2+\sigma)}}(\alpha_1 + \tau\alpha_2 + \tau\alpha_3 + \alpha_4),$$

$$\hat{x}_2 = \frac{1}{(2+\sigma)\sqrt{2}}(\alpha_1 - \sigma\alpha_2 + \sigma\alpha_3 - \alpha_4),$$

$$\hat{x}_3 = \frac{1}{\sqrt{2(2+\tau)}}(\alpha_1 + \sigma\alpha_2 + \sigma\alpha_3 + \alpha_4), \qquad (18)$$

$$\hat{x}_4 = \frac{1}{(2+\tau)\sqrt{2}}(\alpha_1 - \tau\alpha_2 + \tau\alpha_3 - \alpha_4),$$

$$\hat{x}_5 = \frac{1}{\sqrt{5}}(\alpha_1 + 2\alpha_2 + 3\alpha_3 + 4\alpha_4 + 5\alpha_5).$$

The unit vectors $l_i$ can be expressed as a linear combination $l_i = \sum_{j=i}^{5} b_{ij}\hat{x}_j$ where the matrix $B$ is given by

$$B = \frac{1}{\sqrt{10}} \begin{pmatrix} a & \tau & b & -\sigma & \sqrt{2} \\ b & \sigma & -a & -\tau & \sqrt{2} \\ 0 & -2 & 0 & 2 & \sqrt{2} \\ -b & \sigma & a & -\tau & \sqrt{2} \\ -a & \tau & -b & -\sigma & \sqrt{2} \end{pmatrix}, \qquad (19)$$

$$a = \sqrt{2+\tau}, \; b = \sqrt{2+\sigma}, \; \tau = \frac{1+\sqrt{5}}{2}, \; \sigma = \frac{1-\sqrt{5}}{2}.$$

For their relevance to the projection technique of the 5D lattice, we now discuss the projections of only two polytopes $(10000)_{B_5} = \pm l_i \, (i=1,2,...,5)$, and $(00001)_{B_5} = \frac{1}{2}(\pm l_1 \pm l_2 \pm l_3 \pm l_4 \pm l_5)$ into the 3D space described by the symmetry group $W(H_2) \times C_2 \approx D_{5d}$. The other fundamental polytopes lead to some quasiregular polyhedra in 3D space and is not of particular interest here. The group $D_{5d}$ is generated by the group elements $R_1 = r_1 r_3$, $R_2 = r_2 r_4$ and $R_3 = (r_1 r_2 r_3 r_4 r_5)^5$. Here $R_1$ and $R_2$ generate the dihedral group $W(H_2) \approx D_5$ of order 10 and $R_3 = -I$ where $I$ is the $5 \times 5$ unit matrix in the $l_i$ basis. The center of the group $W(B_5)$ is represented by the elements $C_2 = \{I, -I\}$ and therefore it commutes with all the elements of the group. To choose a 3D subspace there are two options; either the space spanned by $(\hat{x}_1, \hat{x}_4, \hat{x}_5)$ or $(\hat{x}_2, \hat{x}_3, \hat{x}_5)$. Let us choose the 3D space defined by the first set of unit vectors. The set of vectors $\pm l_i$ form a single orbit under the group $D_{5d}$. The polytope $(10000)_{B_5} = \pm l_i$ projected into 3D space represents a pentagonal antiprism as shown in Figure 4.

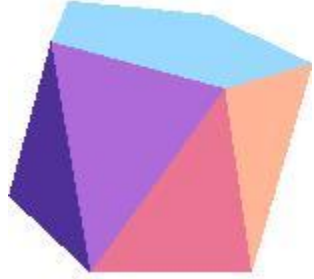

**Figure 4**: The pentagonal antiprism

The polytope $(00001)_{B_5} = \frac{1}{2}(\pm l_1 \pm l_2 \pm l_3 \pm l_4 \pm l_5)$ represents the Voronoi cell of the 5D cubic lattice. The 32 vertices decompose under the group $D_{5d}$ as $32 = 2 + 10 + 20$. The first orbit of size 2 is represents the vectors $\pm \frac{1}{2}(l_1 + l_2 + l_3 + l_4 + l_5)$. Each vector is invariant under the dihedral group $D_5$ but changed to each other under the elements of the center $C_2$. The next orbit of size 10 consists of the vectors like $\pm \frac{1}{2}(-l + l_2 + l_3 + l_4 + l_5)$ with one or four negative signs. They also constitute a pentagonal antiprism like the one in Figure 4. The orbit of size 20 consists of the vectors of with two negative and three negative signs. The union of two orbits 2+20 constitutes a rhombic icosahedron as shown in Figure 5.

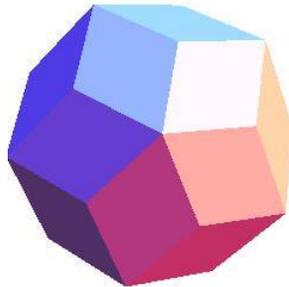

**Figure 5**: The rhombic icosahedron with the symmetry $D_{5d}$

## 3.3 The Coxeter-Weyl group $W(B_6)$ and projections of some of its fundamental polytopes into 3D subspace with $W(H_3) \approx I_h$ symmetry

The icosahedral symmetry is the one which describes some quasicrystal structures in 3D. Here we first discuss how the icosahedral symmetry, the Coxeter group $W(H_3) \approx I_h$, can be obtained as a subgroup of the Coxeter-Weyl group $W(B_6)$. We have already discussed in Section 2 that one of the maximal subgroup of the group $W(B_n)$ is the group $W(D_n)$. Our interest is then the Coxeter-Weyl group $W(D_6)$. We introduce the generators $R_1 = r_1' r_5'$, $R_2 = r_2' r_4'$, and $R_3 = r_3' r_6'$. They generate the Coxeter group $W(H_3)$ [34-36]. Note that in terms of the generators of the group $W(B_6)$ they can be written as $R_1 = r_1 r_5$, $R_2 = r_2 r_4$, and $R_3 = r_3 r_6 r_5 r_6$. They satisfy the relations

$$R_1^2 = R_2^2 = R_3^2 = (R_1 R_3)^2 = (R_1 R_2)^3 = (R_2 R_3)^5 = 1 \tag{20}$$

leading the usual generation relations of the icosahedral group $W(H_3) \approx A_5 \times C_2 = I_h$. The generators of the icosahedral group are $6 \times 6$ matrices in the space of orthonormal vectors $l_i$ and constitute a reducible representation of the icosahedral group. They can be transformed into block-diagonal forms of $3 \times 3$ matrices which act on two sets of orthonormal vectors $(\hat{x}_1, \hat{x}_2, \hat{x}_3)$ and $(\hat{x}_1', \hat{x}_2', \hat{x}_3')$. Each block represents a different $3 \times 3$ irreducible matrix representation.

The block-diagonal form of the generators of the Coxeter group $W(H_3)$ induces the relation

$$\begin{pmatrix} l_1 \\ l_2 \\ l_3 \\ l_4 \\ l_5 \\ l_6 \end{pmatrix} = \frac{1}{\sqrt{2(2+\tau)}} \begin{pmatrix} -1 & -\tau & 0 & -\tau & 1 & 0 \\ 1 & -\tau & 0 & \tau & 1 & 0 \\ 0 & -1 & -\tau & 0 & -\tau & 1 \\ 0 & -1 & \tau & 0 & -\tau & -1 \\ -\tau & 0 & -1 & 1 & 0 & -\tau \\ \tau & 0 & -1 & -1 & 0 & -\tau \end{pmatrix} \begin{pmatrix} \hat{x}_1 \\ \hat{x}_2 \\ \hat{x}_3 \\ \hat{x}_1' \\ \hat{x}_2' \\ \hat{x}_3' \end{pmatrix}. \tag{21}$$

Before proceeding further we note that one can also construct the generators of the Coxeter group $W(H_3)$ as $R_1 = r_1 r_6 r_5 r_6$, $R_2 = r_2 r_4$, and $R_3 = r_3 r_5$ which results from the Dynkin diagram symmetry of $D_6$. Therefore two icosahedral groups are conjugate to each other in the group $Aut(D_6) \approx W(B_6)$.

For projection into 3D space one can use either the first three components or the last three components of the unit vectors $l_i$. Now we discuss the projections of certain $W(B_6)$ polytopes into 3D space with a residual icosahedral symmetry. The polytope with 12 vertices $(100000)_{B_6} = \pm l_i$ are the short roots of the group $W(B_6)$ representing the vertices of a 6D octahedron. The projected copy in 3D space turns out to be an icosahedron represented by the vectors of norm $\frac{1}{\sqrt{2}} \approx 0.707$ as shown in Figure 6.

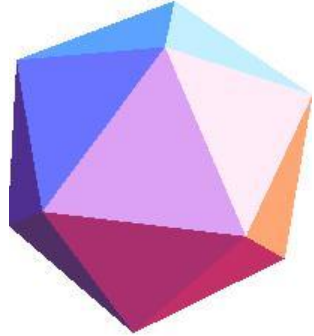

**Figure 6**: An icosahedron projected from the 6D octahedron $(100000)_{B_6}$

The long roots of the group $W(B_6)$ are the vertices of the polytope $(010000)_{B_6}$ which are given by the 60 vectors $(010000)_{B_6} = \pm l_i \pm l_j (i \neq j)$. When they are projected into 3D space they represent two copies of icosidodecahedra with 30 vertices each, one copy is expanded with respect to the other by a factor of $\tau$. Their actual norms in 3D are $\sqrt{\frac{2}{2+\tau}} \approx 0.743$ and $\sqrt{\frac{2}{2+\sigma}} \approx 1.203$. The icosidodecahedron is a polyhedron with 30 vertices, 32 faces (20 triangles+12 pentagons) and 60 edges. One of the icosidodecahedron is depicted in Figure 7.

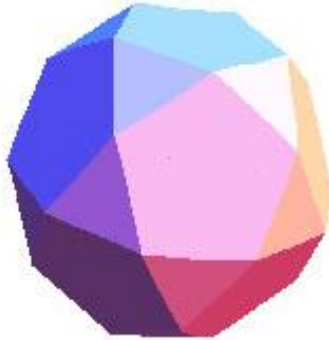

**Figure 7**: Icosidodecahedron projected from the 6D polytope $(010000)_{B_6}$

The polytope $(000001)_{B_6} = \frac{1}{2}(\pm l_1 \pm l_2 \pm l_3 \pm l_4 \pm l_5 \pm l_6)$ represents the Voronoi cell of the 6D cubic lattice with 64 vertices. They decompose into sets with even $(-)$ sign and odd $(-)$ sign representing two orbits of $W(D_6)$ as mentioned in Section 2. We will explicitly demonstrate that each orbit of $W(D_6)$ with 32 vertices decomposes as $32 = 20 + 12$. Normally, the orbits of size 20 and 12 represent a dodecahedron and an icosahedron respectively. However, as we will discuss below, the situation is here such that the dodecahedron (orbit I) and icosahedron (orbit III) form a rhombic triacontahedron, a Catalan solid dual to the icosidodecahedron [32]. The union of the icosahedron (orbit II) and dodecahedron (orbit IV) represents a star dodecahedron. Below we give explicit decomposition of the vertices of 6D cube under the Coxeter group $W(H_3)$. The first two orbits are the decomposition of the vertices with even $(-)$ sign of vectors and the next two are the vertices with odd $(-)$ sign.

orbit I: dodecahedron (I)

$$\pm \frac{1}{2}(l_1+l_2+l_3+l_4+l_5+l_6), \pm \frac{1}{2}(l_1+l_2+l_3-l_4-l_5+l_6),$$

$$\pm \frac{1}{2}(-l_1+l_2+l_3-l_4+l_5+l_6), \pm \frac{1}{2}(-l_1-l_2+l_3-l_4-l_5+l_6),$$

$$\pm \frac{1}{2}(-l_1-l_2-l_3-l_4+l_5+l_6), \pm \frac{1}{2}(l_1+l_2+l_3-l_4+l_5-l_6), \quad (22a)$$

$$\pm \frac{1}{2}(-l_1+l_2+l_3+l_4-l_5+l_6), \pm \frac{1}{2}(l_1-l_2+l_3-l_4+l_5+l_6),$$

$$\pm \frac{1}{2}(-l_1+l_2-l_3-l_4-l_5+l_6), \pm \frac{1}{2}(-l_1-l_2+l_3-l_4+l_5-l_6).$$

orbit II: icosahedron (II)

$$\pm \frac{1}{2}(-l_1-l_2+l_3+l_4+l_5+l_6), \pm \frac{1}{2}(l_1-l_2-l_3-l_4-l_5+l_6),$$

$$\pm \frac{1}{2}(-l_1+l_2-l_3-l_4+l_5-l_6), \pm \frac{1}{2}(-l_1-l_2+l_3+l_4-l_5-l_6), \quad (22b)$$

$$\pm \frac{1}{2}(l_1-l_2-l_3+l_4+l_5+l_6), \pm \frac{1}{2}(-l_1+l_2-l_3+l_4+l_5+l_6).$$

orbit III: icosahedron (III)

$$\pm \frac{1}{2}(l_1+l_2+l_3+l_4+l_5-l_6), \pm \frac{1}{2}(l_1+l_2+l_3+l_4-l_5+l_6),$$

$$\pm \frac{1}{2}(l_1+l_2+l_3-l_4+l_5+l_6), \pm \frac{1}{2}(-l_1+l_2+l_3-l_4-l_5+l_6), \quad (22c)$$

$$\pm \frac{1}{2}(-l_1-l_2+l_3-l_4+l_5+l_6), \pm \frac{1}{2}(l_1-l_2+l_3-l_4+l_5-l_6).$$

orbit IV: dodecahedron (IV)

$$\pm \frac{1}{2}(l-l_2+l_3-l_4-l_5+l_6), \pm \frac{1}{2}(-l_1+l_2-l_3-l_4+l_5+l_6),$$

$$\pm \frac{1}{2}(-l_1-l_2+l_3-l_4-l_5-l_6), \pm \frac{1}{2}(-l_1-l_2-l_3+l_4+l_5+l_6),$$
$$\pm \frac{1}{2}(l_1-l_2-l_3-l_4-l_5-l_6), \pm \frac{1}{2}(-l_1+l_2+l_3-l_4+l_5-l_6), \qquad (22d)$$
$$\pm \frac{1}{2}(-l_1-l_2+l_3+l_4-l_5+l_6), \pm \frac{1}{2}(l_1-l_2-l_3-l_4+l_5+l_6),$$
$$\pm \frac{1}{2}(-l_1+l_2-l_3-l_4-l_5-l_6), \pm \frac{1}{2}(-l_1-l_2+l_3+l_4+l_5-l_6).$$

To see why they represent dodecahedra and icosahedra we just replace the vectors $l_i$ by their first three components in (21). However when we take the union of orbit I and orbit III we obtain the rhombic triacontahedron [32] as shown in Figure 8. The orbit II and the orbit IV form a dodecahedral star which is depicted in Figure 9.

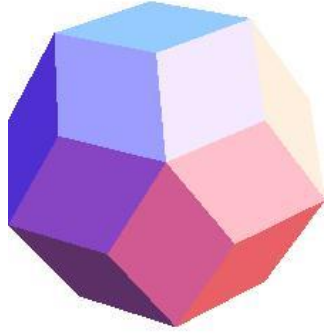

**Figure 8:** The rhombic triacontahedron obtained as part of the projection of 6D cube

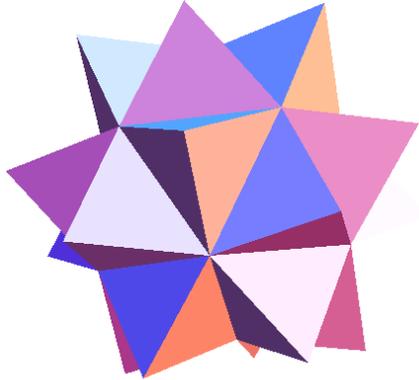

**Figure 9:** The dodecahedral star projected from 6D cube

The vectors representing the vertices of the icosahedra and dodecahedra have the following norms in decreasing order:

$$N(\text{icosahedron III}) = \frac{\tau}{\sqrt{2}} \approx 1.144, \; N(\text{dodecahedron I}) = \sqrt{0.3(2+\tau)} \approx 1.042,$$

$$N(\text{dodecahedron IV}) = \sqrt{0.3(2+\sigma)} \approx 0.644, \; N(\text{icosahedron II}) = \frac{-\sigma}{\sqrt{2}} \approx 0.438.$$

(23)

These are exactly the same results obtained earlier by Conway and Knowles [37] without referring to the overall group structure of the $B_6$ lattice. Note also that

$$N(\text{icosahedron III})/N(\text{icosahedron II}) = \tau^2,$$
$$N(\text{dodecahedron I})/N(\text{dodecahedron IV}) = \tau.$$

(24)

Since the icosahedron obtained from the short roots has a norm 0.707 these three icosahedra follow the ratio: $(1:\tau:\tau^2)$ up to an overall scale factor. We also note in passing that the vectors $\pm l_i$, when projected into 3D subspace, form rhombohedra when taken in groups of three. For example the sets of vectors $(l_1, l_2, l_3)$ and $(l_4, l_5, l_6)$ form an acute and an obtuse rhombohedra respectively as shown in Figure 10.

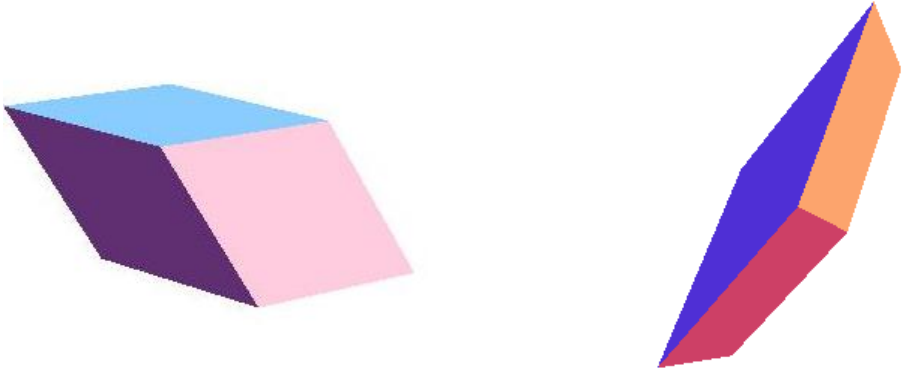

**Figure 10**. The acute and obtuse rhombohedra generated by the vectors $l_i$

## 4. Projections of the lattices of the Coxeter-Weyl groups $W(B_4)$, $W(B_5)$, and $W(B_6)$ into subspaces

The technique which we developed for the projection of the polytopes now can be applied to the lattice points. We will study each case separately. We decompose the $nD$ space into two subspaces $E_{\parallel}$ and $E_{\perp}$ where $E_{\parallel}$ represents the subspace into which the lattice points to be projected and the subspace $E_{\perp}$ is the complementary orthogonal subspace. The shift of the Voronoi cell $V(0)$ along the space $E_{\parallel}$ creates an open strip and the projection of the Voronoi polytope into the subspace $E_{\perp}$ determines a region $K$. It has been shown with a number of examples that the set of lattice points projected from the open strip onto the subspace $E_{\parallel}$ determines a quasicrystallographic structure provided the subspaces admit the noncrystallographic groups as the symmetry groups [38-40]. We will discuss the applications of this procedure in the flowing subsections.

## 4.1 Projection of the lattice points of $B_4$ into a 2D space

We will show here that the projected set of lattice points display a quasicrystal with 8-fold symmetry. The components of a lattice vector $p = a_1\omega_1 + a_2\omega_2 + a_3\omega_3 + 2a_4\omega_4, a_i \in \mathbf{Z}$ along the unit vectors $\hat{x}_i$ are given as follows:

$$p_1 = \frac{1}{2\sqrt{2(2-\sqrt{2+\sqrt{2}})}} (\sqrt{2-\sqrt{2}}a_1 + \sqrt{2}a_2 + \sqrt{2+\sqrt{2}}a_3 + 2a_4)$$

$$p_4 = \frac{1}{2\sqrt{2(2+\sqrt{2+\sqrt{2}})}} (-\sqrt{2-\sqrt{2}}a_1 + \sqrt{2}a_2 - \sqrt{2+\sqrt{2}}a_3 + 2a_4)$$

$$p_2 = \frac{1}{2\sqrt{2(2-\sqrt{2-\sqrt{2}})}} (-\sqrt{2+\sqrt{2}}a_1 - \sqrt{2}a_2 + \sqrt{2-\sqrt{2}}a_3 + 2a_4)$$

$$p_3 = \frac{1}{2\sqrt{2(2+\sqrt{2-\sqrt{2}})}} (\sqrt{2+\sqrt{2}}a_1 - \sqrt{2}a_2 - \sqrt{2-\sqrt{2}}a_3 + 2a_4)$$

(25)

Let us assume that $E_{II} = (\hat{x}_1, \hat{x}_4)$ and $E_\perp = (\hat{x}_2, \hat{x}_3)$. When the Voronoi cell $V(0) = (0001)_{B_4}$ is projected onto the plane $E_\perp = (\hat{x}_2, \hat{x}_3)$ it determines a disc of radius $R_0 = Max(p_2^2 + p_3^2)$. This defines a cylinder in the lattice constraining the integers $a_i$. When the shifted Voronoi cell $V(\omega_4) = (0001)_{B_4} + \omega_4$ is projected into the subspace $E_\perp = (\hat{x}_2, \hat{x}_3)$ the components $(p_1, p_4)$ now determines the quasicrystal structure with 8-fold symmetry as shown in Figure 11. The similar structures are obtained in a recent paper [41]. It consists of squares and rhombi with $45^0$ and the region. The quasicrystal structure with 8-fold symmetry was observed in rapidly solidified $Cr_5Ni_3Si_2$ and $V_{15}Ni_{10}Si$ alloys [42] which can be represented by the quasicrystal structures like in Figure 11.

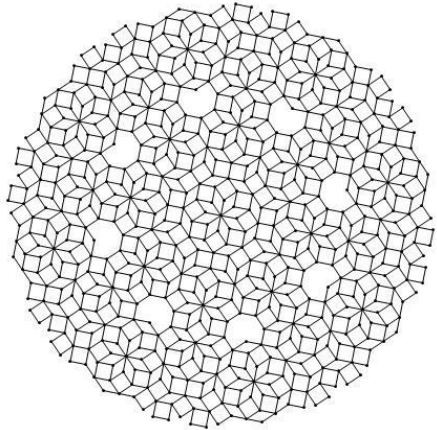

**Figure 11:** The quasicrystal structure obtained from 4D cubic lattice.

## 4.2 Projection of the root lattice of $B_5$ into a 2D subspace

Projection of the 5D cubic lattice onto a 2D plane with 5-fold symmetry has been discussed earlier [21] with no reference to the Coxeter group $W(B_5)$. As we have pointed out in Section 2, the Coxeter number of the group $W(B_5)$ is 10 and has a dihedral subgroup $D_{10}$ of order 20 with a cyclic subgroup of order 10 leading to a 10-fold symmetry. In this section we will study the projections with both symmetries. The 5-fold symmetric projection of the lattice can be obtained by taking the set of orthogonal vectors given by (18). The components of a general root lattice vector $p = (a_1\omega_1 + a_2\omega_2 + a_3\omega_3 + a_4\omega_4 + 2a_5\omega_5)$, $a_i \in \mathbf{Z}$ can be determined by taking the scalar product of the lattice vector with the unit vectors in (18). Projection of the Voronoi cell $V(0)$ into the space defined by the unit vectors $(\hat{x}_2, \hat{x}_3, \hat{x}_5)$ determines the window $K$ which constraints the lattice vectors to be projected onto the plane $(\hat{x}_1, \hat{x}_4)$. When the shifted vector $V(\omega_5) = (00001)_{B_5} + \omega_5$ is projected into the subspace $(\hat{x}_2, \hat{x}_3, \hat{x}_5)$ the distribution of the quasicrystallographic lattice structure in the Coxeter plane $(\hat{x}_1, \hat{x}_4)$ is depicted in Figure 12. The $\hat{x}_5$ direction in this case preserves the translational invariance.

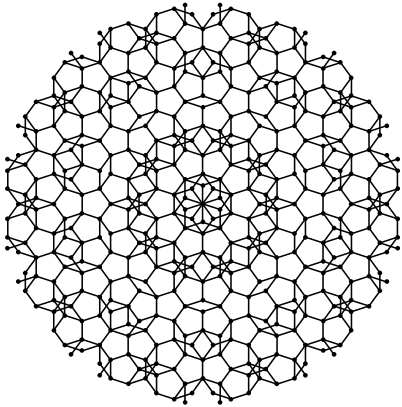

**Figure 12:** 5-fold symmetric quasicrystal structure from projection of the $W(B_5)$ root lattice

The 10-fold symmetric quasicrystal structure by projection of the root lattice of the group $W(B_5)$ can be obtained by using the orthonormal vectors in (10). The unit vectors in (10) follow the sequence of the Coxeter exponents $m_1 = 1, m_2 = 3, m_3 = 5, m_4 = 7, m_5 = 9$ This sequence implies that each of the pair $(\hat{x}_1, \hat{x}_5)$ and $(\hat{x}_2, \hat{x}_4)$ determines one of the Coxeter plane; the first one can be taken as $E_{\parallel}$ and the second pair as $E_{\perp}$ and the unit vector $\hat{x}_3$ defines the direction orthogonal to the planes and preserves the translational invariance. The components of the lattice vectors in the planes and the orthogonal direction are given by

$$E_{II}:\begin{cases} p_1 = \dfrac{1}{\sqrt{10}\sqrt{2-\sqrt{2+\tau}}}\left(-\sigma a_1 + \sqrt{2+\sigma}\,a_2 + \tau a_3 + \sqrt{2+\tau}\,a_4 + 2a_5\right) \\ p_5 = \dfrac{1}{\sqrt{10}\sqrt{2+\sqrt{2+\tau}}}\left(-\sigma a_1 + \sqrt{2+\sigma}\,a_2 + \tau a_3 - \sqrt{2+\tau}\,a_4 + 2a_5\right) \end{cases},$$

$$E_{\perp}:\begin{cases} p_2 = \dfrac{1}{\sqrt{10}\sqrt{2-\sqrt{2+\sigma}}}\left(-\tau a_1 - \sqrt{2+\tau}\,a_2 + \sigma a_3 + \sqrt{2+\sigma}\,a_4 + 2a_5\right) \\ p_4 = \dfrac{1}{\sqrt{10}\sqrt{2+\sqrt{2+\sigma}}}\left(-\tau a_1 + \sqrt{2+\tau}\,a_2 + \sigma a_3 - \sqrt{2+\sigma}\,a_4 + 2a_5\right) \end{cases}$$

$$p_3 = \dfrac{1}{\sqrt{5}}(a_1 - a_3 + a_5)$$

(26)

The quasicrystal structure with 10-fold symmetry from the projection of the root lattice of the Coxeter group $W(B_5)$ is shown in Figure 13.

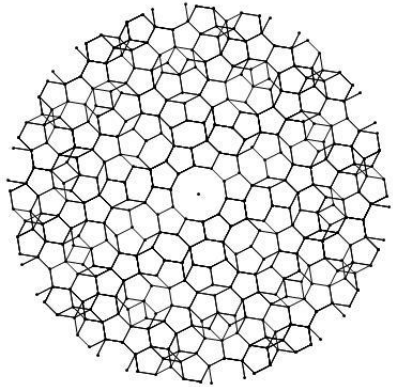

**Figure 13:** 10-fold symmetric quasicrystal structure from 5D cubic lattice

### 4.3 Projection of the root lattice of $B_6$ into 2D and 3D subspaces

The Coxeter number of the group $W(B_6)$ is 12. Therefore it is quite natural to expect a 12-fold symmetric quasicrystal structure in a plane from the projection of the root lattice of the 6D cubic lattice. We have studied projection of the 6D cubic lattice in another publication [19-20] using the formula (10) for the group $W(B_6)$ and obtained the following quasicrystal structure in Figure 14.

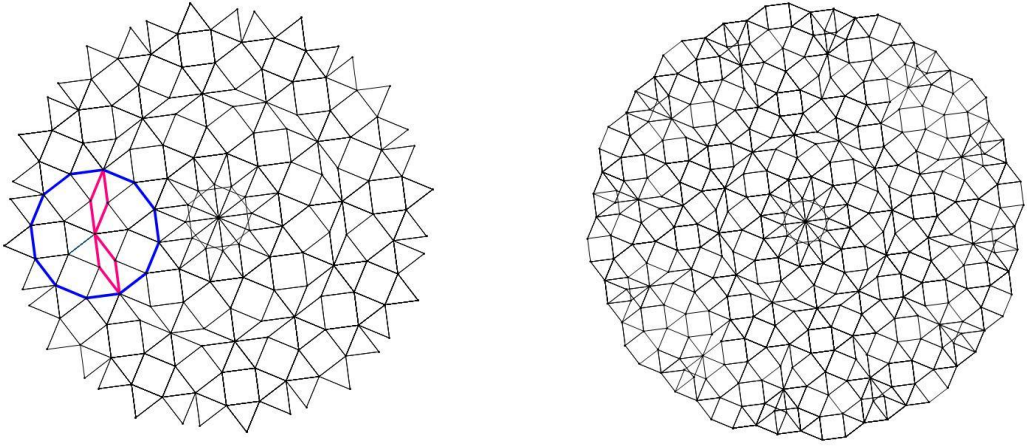

**Figure 14:** 12-fold symmetric quasicrystal structure from 6D cubic lattice

The Figure 14 shows that the dodecagonal tiling displayed here is different than the usual tiling which is based only on the triangle and square tiles. It also involves rhombi in addition square and triangle tiles. Such a structure has been recently observed in the dodecagonal quasicrystal formation of $BaTiO_3$ [43].

The projection of the 6D cubic lattice into 3D subspace has been discussed in various papers earlier without taking into account the detailed structure of the Coxeter group $W(B_6)$ [12, 37, 39]. In Section 3.3 we studied explicitly the icosahedral subgroup $W(H_3)$ of the Coxeter group $W(B_6)$ and showed that the space can be decomposed as $E_{II}$ and $E_\perp$ defined by the unit vectors $(\hat{x}_1, \hat{x}_2, \hat{x}_3)$ and $(\hat{x}'_1, \hat{x}'_2, \hat{x}'_3)$ respectively. A general root lattice vector can be written as

$$p = \frac{1}{\sqrt{2(2+\tau)}}[(n_1+n_2\tau)\hat{x}_1 + (n_3+n_4\tau)\hat{x}_2 + (n_5+n_6\tau)\hat{x}_3 + \tau(n_1+n_2\sigma)\hat{x}'_1 + \tau(n_3+n_4\sigma)\hat{x}'_2 + \tau(n_5+n_6\sigma)\hat{x}'_3], \quad (27)$$

where the integers $n_i$ are given by

$$n_1 = -a_1, n_2 = -a_5, n_3 = -(a_3+2a_4+2a_5+2a_6),$$
$$n_4 = -(a_1+2a_2+2a_3+2a_4+2a_5+2a_6), n_5 = -(a_5+2a_6), n_6 = -a_3. \quad (28)$$

A similar formula to (27) involving just the coefficients of the unit vectors $\hat{x}_1, \hat{x}_2$, and $\hat{x}_3$ were obtained earlier from another consideration [44].

The root vectors $\beta_i$ $(i = 1,2,3)$ of the Coxeter diagram of $W(H_3)$ can be determined as $\beta_1 = -\sqrt{2}\hat{x}_1, \beta_2 = \frac{1}{\sqrt{2}}(\hat{x}_1+\sigma\hat{x}_2+\tau\hat{x}_3), \beta_3 = -\sqrt{2}\hat{x}_3$. Note that the pairs of vectors $(\beta_1,\beta_3), (\beta_1,\beta_2)$, and $(\beta_2,\beta_3)$ determine the 2-fold, 3-fold, and 5-fold symmetry planes respectively. By choosing suitable orthogonal vectors in these planes such as $(\hat{x}_1,\hat{x}_3), (\hat{y}_1,\hat{y}_2)$, and $(\hat{z}_1,\hat{z}_2)$ where

$$\hat{y}_1 = \frac{1}{2}(-\hat{x}_1 + \sigma\hat{x}_2 + \tau\hat{x}_3), \hat{y}_2 = \frac{-1}{2\sqrt{3}}(3\hat{x}_1 + \sigma\hat{x}_2 + \tau\hat{x}_3),$$
$$\hat{z}_1 = \frac{1}{2}(\tau\hat{x}_1 - \hat{x}_2 + \sigma\hat{x}_3), \hat{z}_2 = \frac{1}{2\sqrt{2+\tau}}(\hat{x}_1 + \sigma\hat{x}_2 + (2+\tau)\hat{x}_3),$$
(29)

one can project the lattice vectors in (27) onto these planes provided the $E_\perp$ components of the vectors remain in the window **K** determined by the projection of the Voronoi cell of the 6D cube.

The distribution of the lattice points are displayed in Figure 15, Figure 16, and Figure 17.

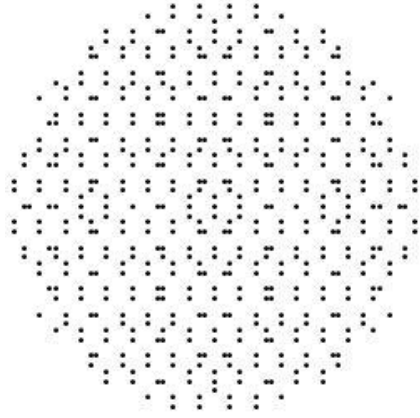

**Figure 15.** 2-fold symmetry from projection of $B_6$ lattice

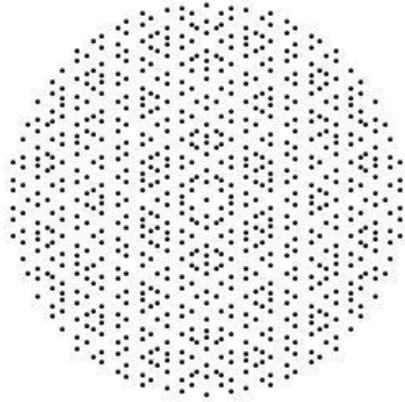

**Figure 16.** 3-fold symmetry from projection of $B_6$ lattice

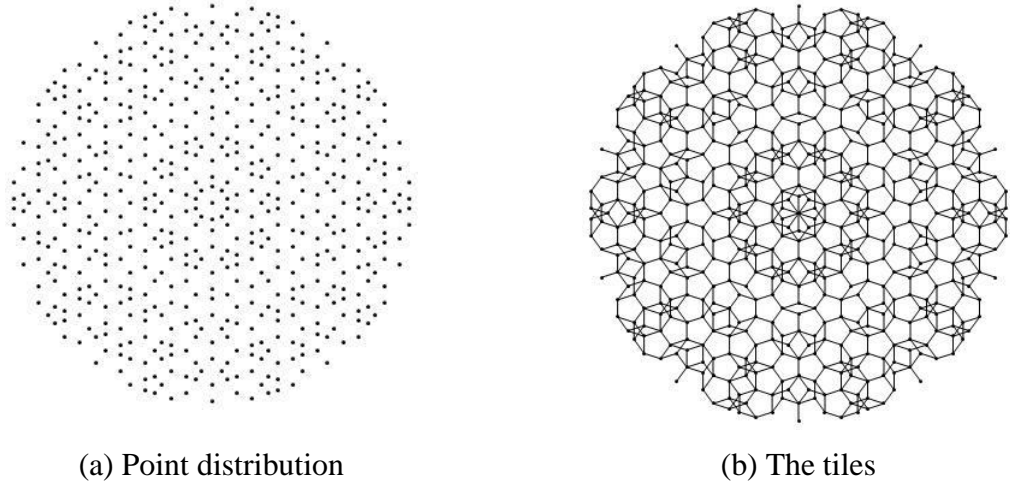

(a) Point distribution             (b) The tiles

**Figure 17:** 5- fold symmetry from projection of $B_6$ lattice

## 5. Conclusions

We presented a systematic analysis of the higher dimensional lattice projection technique with an emphasis on the group theoretical structure of the *n*D cubic lattices. It is proposed that the Coxeter number $h$ of the Coxeter-Weyl group $W(B_n)$ plays a crucial role in the determination of the dihedral subgroup $D_h$ of the group $W(B_n)$ that in turn determines the symmetry of the quasicrystal structure in the Coxeter plane. The eigenvalues and eigenvectors of the Cartan matrix (Gramm matrix) lead to the correct choice of the Coxeter plane onto which the lattice is projected. We reproduce the earlier results obtained from the lattice projections. In addition, in two cases we obtained new results. The 10-fold symmetric quasicrystallography from $W(B_5)$ and the 12-fold symmetric quasicrystal structure from $W(B_6)$ lead to some novel structures. In particular the tiling displayed in Figure 14 is compatible with a recent experiment with 12-fold symmetry [43]. The technique developed here can be extended to any higher dimensional lattice described by the whole series of affine Coxeter groups. If the argument raised here is verified by the experimental quasicrystallography then the Coxeter numbers and Coxeter exponents will gain some status in nature.